# Magnetically Localizing Heat or Enhancing Equilibration in a Quasi 1D Magnetic Fluid

Jun Huang and Weili Luo[1]

Department of Physics, University of Central Florida, Orlando, FL 32816

Tianshu Liu

Department of Mechanical and Aerospace Engineering, Western Michigan University

Kalamazoo MI 49008-5343

Abstract

Using two different configurations of temperature and magnetic field gradients, we observed that, in a quasi one-dimensional magnetic fluid, magnetic force either reduces the temperature difference across the sample when the two gradients are parallel to each other (PL), or increase the temperature difference when the two gradients are antiparallel (AP), where the single convection roll in zero field was replaced by two localized flows at the two ends of the sample cell. This flow structure stops the heat flow of approaching to thermal equilibrium in the system, causing the temperature at hot side of the sample cell getting hotter and cold side becoming colder. None of these phenomena can be described by the existing theories of magnetically-induced instabilities. The underlying physics for observed results for AP configuration has been proposed as the mechanism to drive a new type of heat engines that has much higher efficiency than Carnot engines and has no pollution to the environment, our results point to the potential feasibility of this proposed mechanism.

[1] weili.luo@ucf.edu

An external force can change the way energy is transferred in a fluid system. For example, in a gravitational field gravito-thermal convection occurs when a fluid is heated from the bottom and a critical temperature difference across the horizontal layer of a fluid is reached. An applied magnetic force can drive a fluid to convective instability similar to gravity. Lalas, Carmi, and Curtis proposed that a uniform magnetic field gradient behaves like gravity and can drive a layer of magnetic fluid in a temperature gradient to a convective instability [1-2]. Finlayson [3] proposed in 1970 that if a vertical uniform magnetic field is applied on a magnetic fluid (MF) along a temperature gradient, convective instability can result from the temperature-dependent magnetization, which gives rise to a temperature-dependent internal field whose gradient is a function of temperature gradient. Several groups have observed field-induced instability, most of them in a uniform magnetic field [4-9].

In this paper we report the observation of the magnetically-induced localization of heat in a quasi 1D magnetic fluid that supressed convective instability and drove the system to an increasingly more stable state with increasing field where the heat energies with different temperatures were separated. This result can not be described by the field-induced convective instabilities in Ref [1-3]. Because the underlying physics was proposed as foundation for a new type of heat engines that is drastically different from Carnot Engine with much higher efficiency [10], we will discuss the implication of this work at the end of this paper. Additionally, filed-induced non-localized flow was also observed in a second configuration that was used as a control experiment, which seems also beyond the description by Finlayson's mechanism.

The schematic of the experimental set up is shown in Fig. 1. The two exactly matched sample cells with separation of 12.2 cm were arranged side by side inside two electromagnetic poles; each was enclosed inside a vacuum chamber. Both were heated on the left side with electric heaters and were cooled from right side with running coolant from a cold tank so that the temperature gradients in both cells were towards left. The magnetic field and the field gradient were designed such that the magnetic force was perpendicular to the gravity and the spatial distribution of the field was symmetric about the centre. The field gradients were toward the magnetic poles for the two cells so that for the left cell the gradients of temperature and the applied field were parallel (PL), for the right cell antiparallel (AP), to each other. The sample is a quasi 1D magnetic fluid (MF) consisting magnetic nanoparticles of 10 nm suspended in nonmagnetic solvent with volume fraction of 1% [11-14]. For our dilute concentration, we can consider that the fluid exhibits super-paramagnetic behaviour. The sample cell in our experiment is made of Acrylic with the dimension 90mmx7.5mm x5.5mm with its long axis, the temperature gradient, applied magnetic field, and applied field gradient all in the horizontal direction that is perpendicular to the gravity. When horizontal heating and cooling are applied in zero field, the buoyancy force due to thermal expansion drives the system to convective instability.

In Fig. 2 we plot the temperature as function of time for different applied fields obtained from the thermocouples at the four corners of each sample cell, with Fig. 2(a) for parallel (PL) and Fig 2 (b) for antiparallel (AP) configurations, respectively. The insets in both Fig. 2 (a) and Fig. 2 (b) show the labels for the thermocouple positions: LLT denotes one for left cell (PL), left side (hot side), and the top positon; LRB for left cell, right side (cold side), and bottom location; RRT for right cell (AP), right side (cold side), and top location, etc. Once the heating and cooling were applied at t = 0 sec, the convective flow started [12]. The temperatures in the two samples reached a quasi-steady states at t ≥ 1000 sec. in zero field. Then the magnetic field was applied by steps: at t = 2000s, 10mT was applied; we waited for 200 sec. in this field; then 20mT was applied for 200sec., then 30mT, etc. At each step the field was increased by 10 mT. The field values here refer to the ones measured at the magnet poles. Therefore, they are the maximum values for that applied field along the axis of the two poles. For PL configuration, applying field reduces the temperature difference across the sample cell (from LLT to LRT; LLB to LRB) as well as from top to bottom of the cell (LLT to LLB and LRT to LRB) as shown in Fig. 2 (a); for AP configuration, the applied field enhances the temperature difference across the sample (RLT to RRT; RLB to RRB), clearly indicating that the applied field could drastically alter the convective heat transfer, depending on the configuration of the experiment. Fig. 3 shows the typical temperature fields in the base flow with no field, for PL and AP configurations in field of $B_{max}$ = 80mT. Clearly, the high and low temperature regions were isolated by the field applied in AP situation, while the field applied in PL arrangement facilitated heat transfer from the hot end to the cold end.

To understand the underlying physical mechanism for these different field-dependent temperature changes, we need to know flow structures as function of field. For the *opaque (black)* MF that is difficult for flow visualization, we have adopted a two-step image-processing method to extract high-resolution velocity fields in a cell. In the first step, velocity fields were extracted from a sequence of images of microspheres particles in a kerosene fluid (as the transparent base fluid) by using the particle image velocimetry (PIV) with the optical flow method. Cenosphere particles were added to the base fluid as trace particles [15]. Since the concentration of the particles is only 1% in volume fraction, it is a reasonable approximation to assume that the flow structures in *zero field* for MF is essentially the same as that of base fluid. The time-averaged velocity field was obtained as the base flow velocity field when a quasi-steady state of the flow was achieved in a sufficient time after the heating and cooling were turned on. The streamlines of the base flow field in zero field is plotted in Fig. 4 as a reference for both PL and AP cases. A single convective roll is clearly shown along with some finer vortices generated by the Kevin-Helmholtz instabilities [16] at the interface between the two opposite flows along the horizontal direction.

Temperature-sensitive paint (TSP) was used to obtain high-resolution temperature fields on the cell surface in the heating and cooling conditions before a magnetic field is applied [17], where temperature profiles were obtained by florescent light intensity from TSP on sample surface. The second step is to extract the perturbation velocity field of MF from TSP images in the same heating and cooling conditions after the field is applied. Then, the velocity field is reconstructed by superposing the base flow velocity field and the perturbation velocity field. We consider the temperature decomposition $\theta = \langle \theta \rangle_T + \theta'$ and the velocity decomposition $\bm{u} = \langle \bm{u} \rangle_T + \bm{u}'$, where $\langle \bullet \rangle_T$ is the time-averaging operator, $\theta'$ and $\bm{u}'$ denote the time-dependent perturbation terms, $\langle \bm{u} \rangle_T$ is the base flow velocity field, and $\langle \theta \rangle_T$ is the time-averaged temperature field before the field is applied. The time-averaged quantities satisfy the steady-state energy transport equation. The equation for the perturbations is $\frac{\partial \theta'}{\partial t} + \nabla \cdot (\bm{u}' \langle \theta \rangle_T) = f$ where $f$ is a term related to the perturbation correlations. This equation has a similar form of the physics-based optical flow equation [18]. The perturbation velocity $\bm{u}'$ was obtained from the time-dependent temperature change associated with the field applied to MF by solving the optical flow problem where $f = 0$ is assumed in the first-order approximation. Then, the complete velocity field of MF is reconstructed by a superposition of the base flow velocity and the perturbation velocity for the magnetic field at certain time.

In Figs. 4(a) and 4(b), the streamlines are plotted for PL and AP configurations, respectively, in the applied fields of $B_{max}$ = 20mT, 40mT, 60mT, and 80mT. Even for low field strength, the flow structure has changed considerably in comparison to that in zero field. For PL geometry, the convective boundary-layer flow in zero field has been replaced by a large vortex whose centre shifts towards the left end (higher field and higher temperature) with increasing field. In contrast, for AP arrangement, the applied field breaks the circulating flow in zero field into two localized flow structures near the two ends of the sample cell, indicating the qualitative flow-topological change induced by applied field. There is a distinct saddle point between the two localized flow structures. The flow structures, from a fluid-dynamic perspective, explain the observation that the magnetic field increases the temperature difference between the hot and cold ends in AP configuration as shown in Fig. 3. This is because the localized flow structures interrupt the convective roll that carries the heat from the hot to the cold end and vice versa.

Convective instability induced in a magnetic field gradient in MF proposed by Lalas, Carmi, and Curtis [1-2] does not have to be in the vertical, i.e. the gravity, direction. In a horizontal direction, it can occur as well. For this instability to occur, however, the fluid has to be in a potentially unstable situation. The unstable configuration involves the non-uniform magnetic field applied to a magnetic fluid layer that has a temperature gradient, where field gradient is parallel to the temperature gradient but antiparallel to the magnetization gradient. The magnetization at colder temperature is larger than that at higher temperature and experiences a magnetic force that is along the field gradient direction, leading to magnetically induced convective instability. In our experiment, however, neither PL nor AP configuration is potentially unstable. For AP case, both lowest temperature and highest field are closest to the pole at the right where the magnetization has the maximum value. The magnetic body force also points to the right direction, driving fluid elements with high magnetization towards to the high field and lower temperature. Therefore, AP configuration is very stable against the so-called instability. Furthermore, from the streamlines in Fig.4 (b), the localized flow induced by the magnetic force inhibits the gravito-thermal convection. As the result the communication for the heat energy between the hot and cold sides was suppressed. For PL case, magnetic force points to the left pole. Although the temperature at the end closest to the left pole is higher than the lower field side, the magnetization value is maximum due to the proximity to the pole. Again, this is not the situation conducive to the magnetically-induced convective instability by a field gradient. In Finlayson's theory the applied field is uniform, therefore, the applied magnetic force is zero. Due to the temperature-dependent magnetization, the internal field is not uniform and its gradient, i.e. the internal magnetic force, is a function of temperature gradient. In our experiment, the contribution to the magnetic body force from temperature gradient is dominated by the applied field gradient, therefore, Finlayson's theory can not describe our observation, either. From mathematics point of view, the global vortex structure appeared in the PL configuration and localized flows in the AP configuration are the two solutions from a bifurcation point as the result of the special geometry of the sample, the specific configuration of the field and temperature gradients that warrant theoretical studies in the future.

We plot in Fig 5 (a) the vertical velocity component averaged across the vertical coordinate, $U_y$, versus a specific location along the horizontal direction, x, for PL and AP geometries in zero and applied fields. The increment of field is 20 mT. Fig 5(b) are the similar plots but for horizontal velocity component averaged across the vertical coordinate, $U_x$, versus x. For AP case, because of the heat energy in the system is not communicated between the hot and cold sides, part of the field-induced localized thermal energy was transferred to kinetic energy, whose values are much larger than that of PL case at the corresponding ends for both $U_y$ and $U_x$. For both configurations, the increases in magnitudes of $U_y$ and $U_x$ are larger at the ends of the sample cell, especially at the sides where the field and field gradient are smallest. Because the magnetization are the largest closest to the poles and the smallest at the ends furthermost from the poles, we believe that higher fields at the poles lead to larger viscosity that impede the flow motion, which explains the relative small magnitudes of the velocities $U_y$ and $U_x$ for both PL and AP cases at the higher-field ends.

This experiment was conducted in order to verify the principle for a new type of Non-Carnot Heat Engine proposed by one of us [10], which has higher efficiency than Carnot engine and it has not pollution to the environment. The idea behind it was if the configuration of the magnetic field gradient and the temperature gradient are anti-parallel to each other, as in our AP geometry, the heat flow from high to low temperature will be halted and the overall temperature difference across the sample will increase with increasing magnetic fields (and the field gradients), as the case in our AP geometry. The condition for this to happen is that the driving force is point to lower temperature and higher field for AP geometry. We have calculated magnetic body force with experimental parameters and found that it satisfies this requirement [19].

In summary, two drastically different types of temperature changes across the sample as a function of field were observed in our quasi one-dimensional magnetic fluid sample, depending on the relative orientation of temperature gradient *vs*. field gradient. The discrepancy originates from the dissimilar flow structures induced in field. Neither results from AP and PL configurations can be described by the field-induced instabilities in the existing theories. Our results for AP configuration confirmed the validity of the proposed mechanism for a new type of Non-Carnot heat engine, pointing to the feasibility of the new technology.

In our experimental time scale the effect due to magneto-diffusive convection is not important due to the long time needed to establish the mass gradients [20-23].

Acknowledgement

We thank Dr. Kuldip Raj from Ferrotech for providing the sample MF 905 used in this experiment.

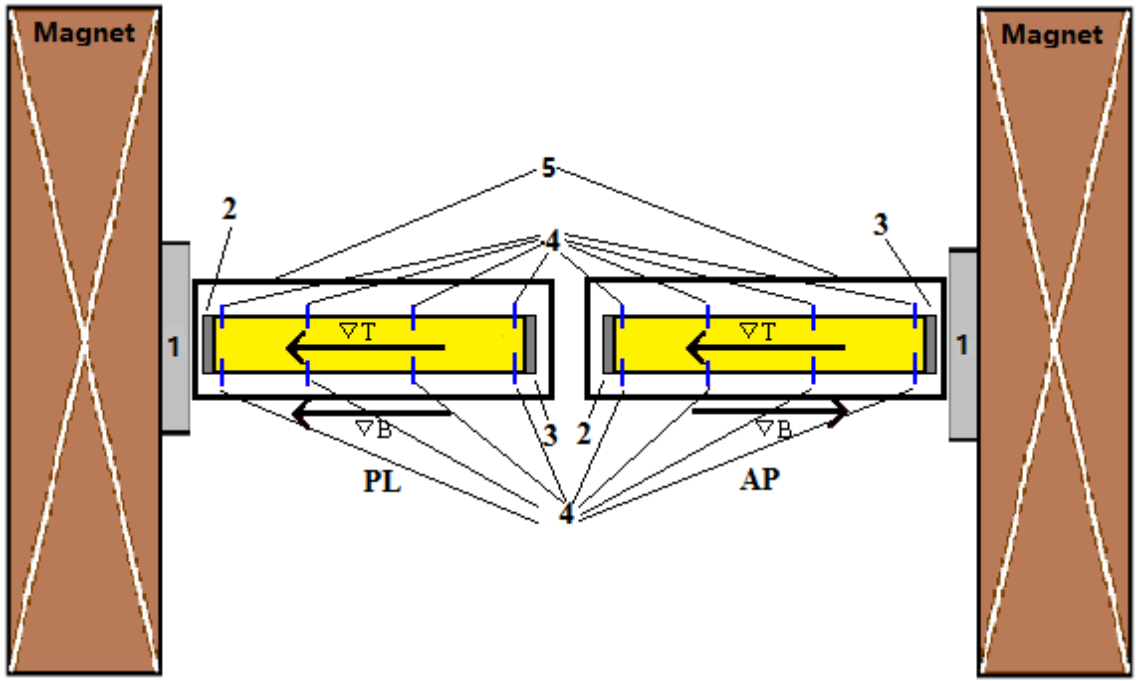

Fig. 1. Experimental Set up. 1: magnet poles; 2: electric heaters for both cells; 3: the cooling fluid running through the right sides of the both cells; 4: 8 thermocouples for each sample cell to monitor the temperature at top and bottom of the cell; 5: vacuum chambers for both cells. For the left cell, the gradients of temperature and field are parallel to each other (PL), and for the right cell, antiparallel to each other (AP).

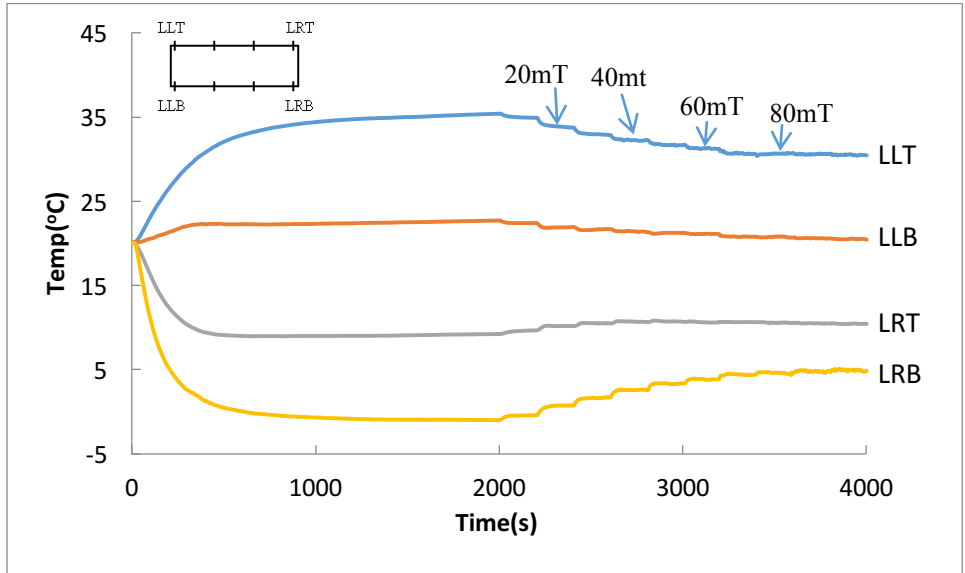

(a)

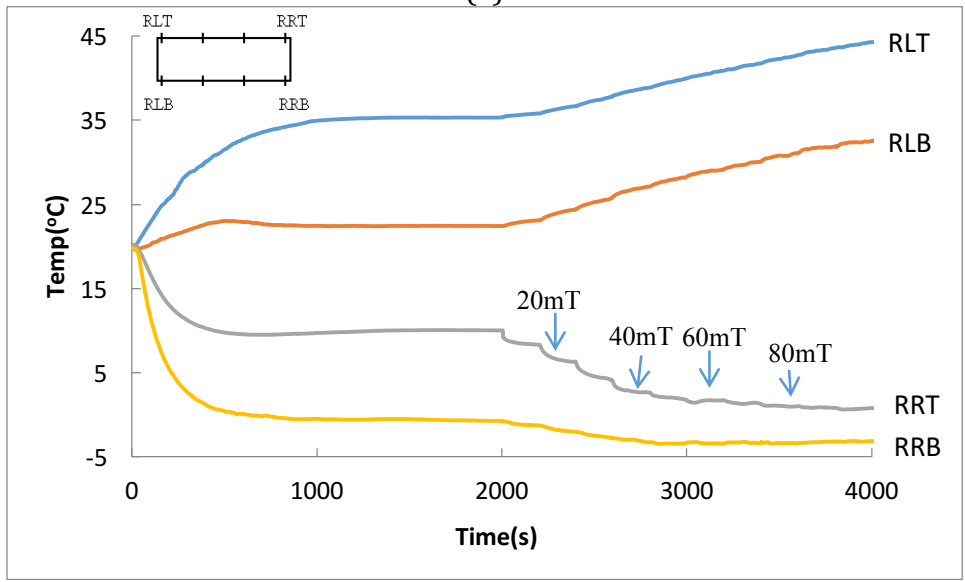

(b)

Fig 2. Temperature measured by thermal couples at the four corners of each sample cell as a function of time for (a) parallel (PL) and (b) antiparallel (AP) configurations in zero field from 0-2000 sec. and their changes when fields were applied afterward. The insets show the labels for positions of the four thermal couples in each cell.

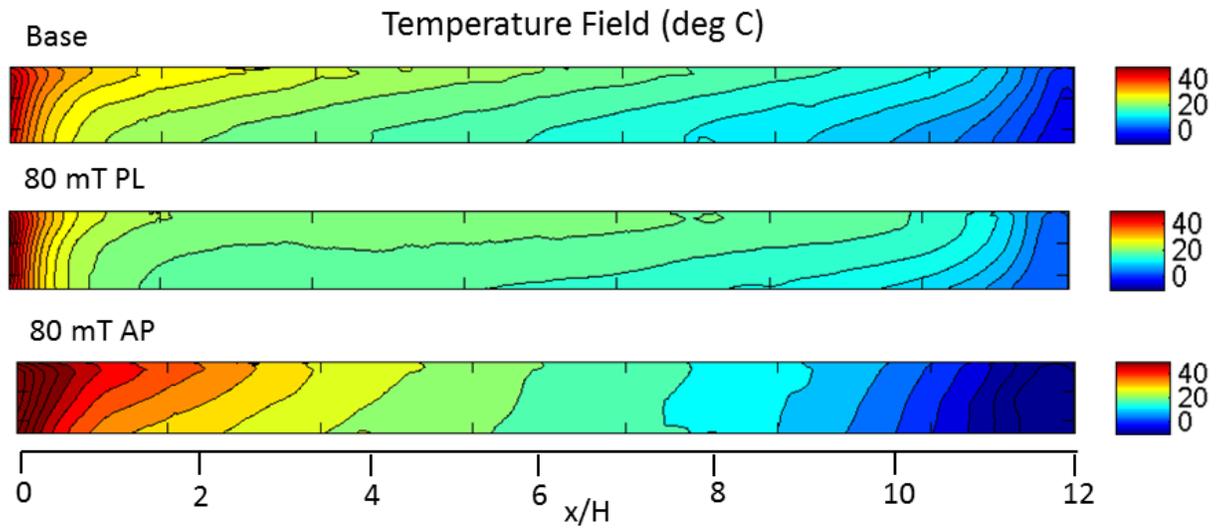

Fig 3. Temperature fields in the base flow with no field, PL ($B_{max}$ = 80mT), and AP ($B_{max}$ = 80mT) configurations. The horizontal axis is scaled by the vertical height of the sample cell, H.

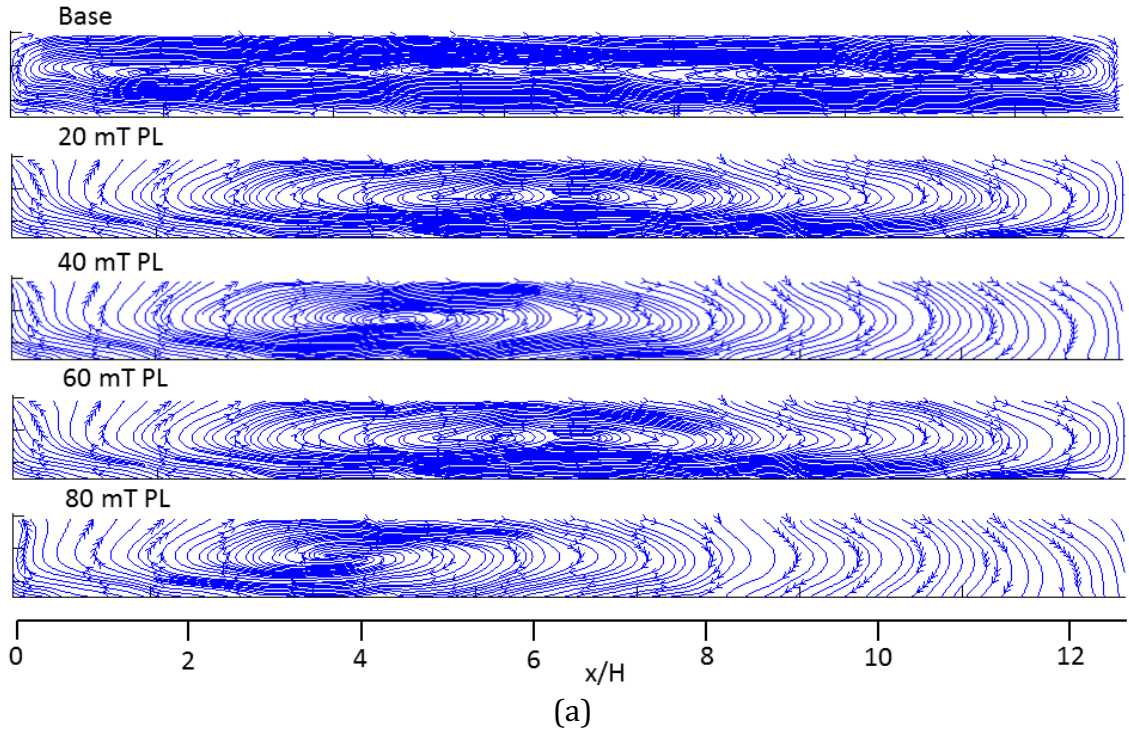

(a)

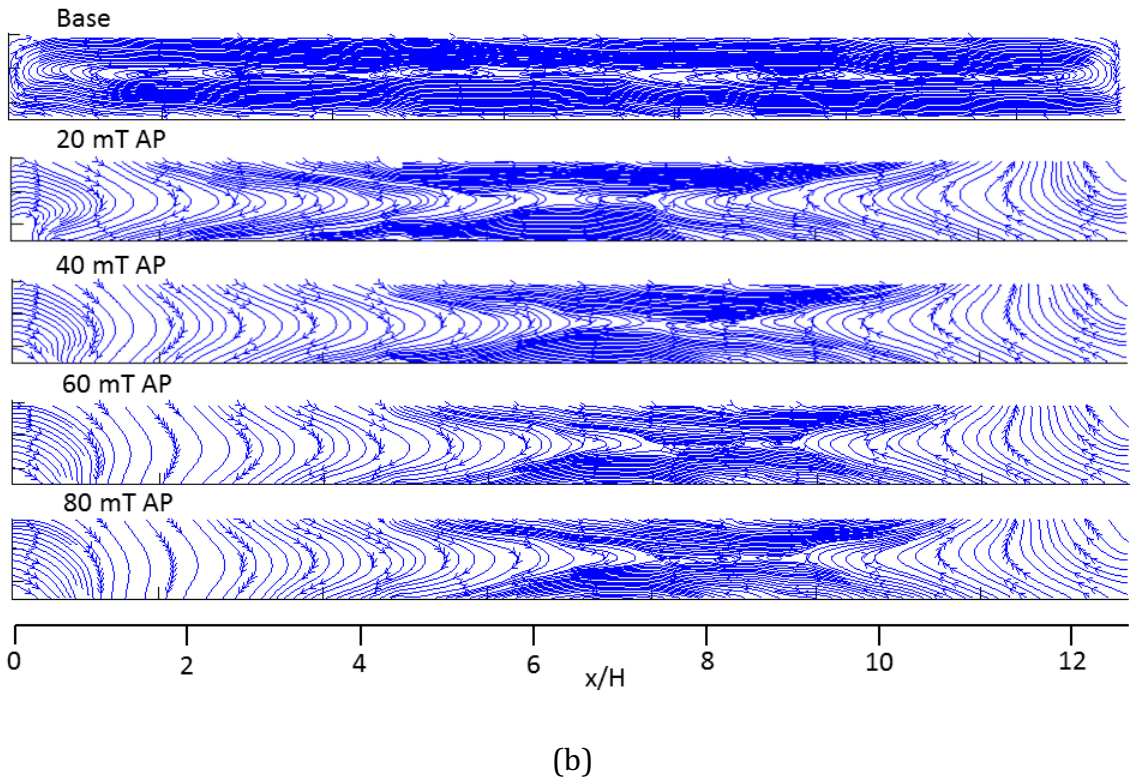

(b)

Fig 4. Streamline patterns for (a) PL and (b) AP configurations in fields of $B_{max}$ = 20mT, 40mT, 60mT, and 80mT. The field values here referring to the maximum values at the poles. The horizontal axis is scaled by the vertical height of the sample cell, H.

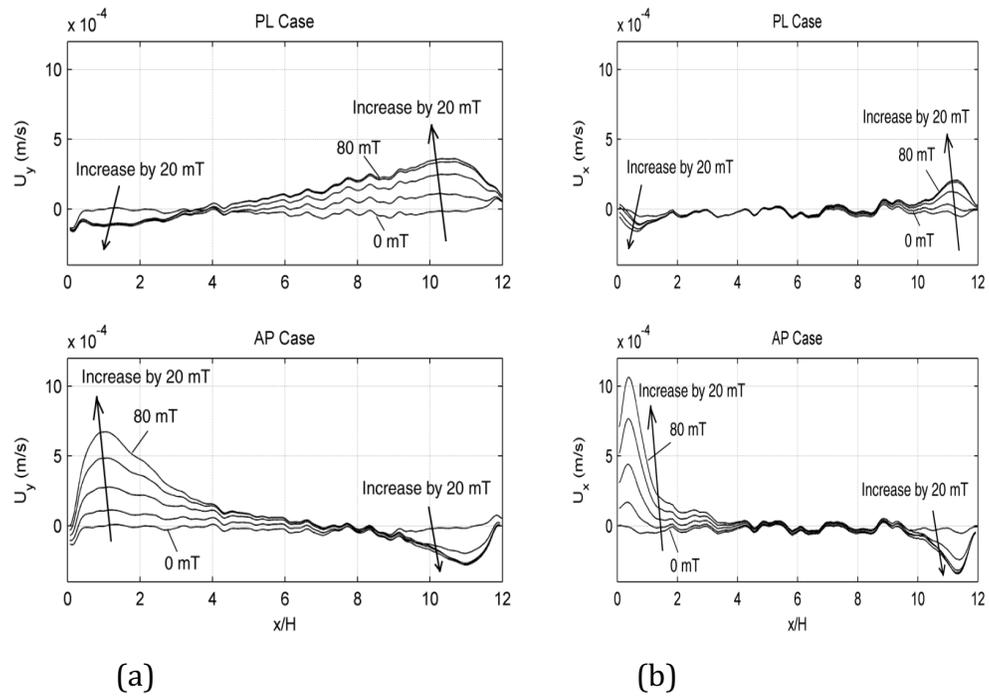

(a)                                              (b)

Fig. 5. The profiles of (a) vertical and (b) horizontal velocity components averaged across the vertical coordinate for parallel and antiparallel configurations. The horizontal axis is scaled by the vertical height of the sample cell, H.